\documentclass[twocolumn,showpacs,amsmath,amssymb,prb]{revtex4}

\usepackage{graphicx}
\usepackage{dcolumn}
\usepackage{bm}
\usepackage{xcolor}

\begin{document}

\preprint{APS/123-QED}
\title{Nonsaturating magnetoresistance, anomalous Hall effect, and magnetic quantum oscillations in ferromagnetic semimetal PrAlSi}
\author{$^{1,2}$Meng Lyu}
\author{$^{1}$Junsen Xiang}
\author{$^{1}$Zhenyu Mi}
\author{$^{1,2}$Hengcan Zhao}
\author{$^{1,2}$Zhen Wang}
\author{$^{1,2,3}$Enke Liu}
\author{$^{1,2,3}$Genfu Chen}
\author{$^{1,2,3}$Zhian Ren}
\author{$^{1,2,3}$Gang Li}
\author{$^{1,2,3}$Peijie Sun}
\affiliation{%
$^{1}$Beijing National Laboratory for Condensed Matter Physics, Institute of Physics, Chinese Academy of Sciences, Beijing 100190, China\\
$^{2}$University of Chinese Academy of Sciences, Beijing 100049, China \\
$^{3}$Songshan Lake Materials Laboratory, Dongguan, Guangdong 523808, China \\
}

\date{\today}

\begin{abstract}
We report a comprehensive investigation of the structural, magnetic, transport and thermodynamic properties of a single crystal PrAlSi, in comparison to its nonmagnetic analogue LaAlSi. PrAlSi exhibits a ferromagnetic transition at $T_{C}$ = 17.8 K which, however, is followed by two weak phase transitions at lower temperatures. Based on the combined dc and ac magnetic susceptibility measurements, we propose the two reentrant magnetic phases below $T_C$ to be spin glasses or ferromagnetic cluster glasses. When the magnetic glassy states are suppressed by small field, several remarkable features appear. These include a linear, nonsaturating magnetoresistance as a function of field that is reminiscent of a topological or charge-compensated semimetal, and a large anomalous Hall conductivity amounting to $\sim$2000 $\Omega ^{-1}$cm$^{-1}$. Specific-heat measurements indicate a non-Kramers doublet ground state and a relatively low crystal electric field splitting of the Pr$^{3+}$ multiplets of less than 100 K. Shubnikov-de Hass oscillations are absent in LaAlSi, whereas they are clearly observed below about 25 K in PrAlSi, with an unusual temperature dependence of the dominating oscillation frequency $F$. It increases from $F$ = 18 T at 25 K to $F$ = 33 T at 2 K, hinting at an emerging Fermi pocket upon cooling into the ordered phase. These results suggest that PrAlSi is a new system where a small Fermi pocket of likely relativistic fermions is strongly coupled to magnetism. Whether hybridization between $f$ and conduction band is also involved remains an intriguing open problem.
\end{abstract}

\pacs{Valid PACS appear here}
\maketitle

\section{introduction}
Weyl nodes are linear band touching points of electron and hole pockets with distinct chirality in three-dimensional Brillouin zone \cite{armi18}. While they were first predicted to exist in systems with broken time-reversal symmetry \cite{wan11,xu11}, the subsequent discovery of nonmagnetic Weyl semimetals, such as TaAs, without space-inversion symmetry has attracted tremendous attention \cite{weng15}. Since then, searching for topological semimetals with broken symmetries of both types also develops into an important research field \cite{liu17}. Here, the coexistence of broken space-inversion and time-reversal symmetries provides an opportunity to study the interplay between relativistic fermions and internal magnetism of a magnetic topological semimetal \cite{burkov15}. Best-recogonized examples of such interplay include the large anomalous Hall effect (AHE) and anomalous Nernst effect arising from the enhanced Berry curvature  \cite{enke18,qwang18,sakai18}. On the other hand, magnetic semimetals are interesting on themselves because of, for example, the spin-dependent electronic states and charge transport \cite{zhang08}, which may lead to potential applications in spintronics \cite{smej17}. Finally, we focus on magnetic semimetals also because they are frequently related to Mott or Kondo physics, suited for exploring electron-correlated behaviors. Such cases are probably realized in the correlated Dirac semimetal of perovskite CaIrO$_3$ \cite{fujioka19} and the Weyl Kondo semimetals YbPtBi \cite{guo18} and CeRu$_4$Sn$_6$ \cite{dai17}.

Recently, rare-earth based compounds $R$Al$X$ ($R$ = Ce and Pr; $X$ = Si and Ge) attract increasing attention as potential candidates of magnetic Weyl semimetals \cite{chang18,hodo18,pub19,meng19}. They may crystallize in two alternative body-centered tetragonal lattices \cite{guloy91}. One has space group $I4_1$$md$ (No. 109) and is derived from the prototype LaPtSi (ref. \cite{sharma07}). This type of structure is noncentrosymmetric and has a polar point group $4mm$. The other is derived from $\alpha$-ThSi$_2$ of $I$4$_1$/$amd$ space group (No. 141) that has a centrosymmetric point symmetry of $4/mmm$. In contrast to the distinct occupations of different $4a$ Wyckoff sites by Al and $X$ atoms in the former, they occupy the $8e$ site randomly in the latter case \cite{flan98}. A detailed comparison of the two structures has been made in a recent paper \cite{hodo18}. Among the $R$Al$X$ family, compounds with $X$ = Ge have been relatively well investigated thus far, albeit with considerable controversies on the magnetic properties. For example, while $R$AlGe was theoretically predicted to be a ferromagnetic (FM) Weyl semimetal \cite{chang18}, experimentally CeAlGe was found to be either antiferromagnetic (AFM) \cite{hodo18, pub19} or FM \cite{flan98}, and PrAlGe either a spin glass \cite{pub19} or a ferromagnet \cite{meng19}. On the other hand, the silicide $R$Al$_x$Si$_{2-x}$ with $x$ $\sim$ 1 has also been investigated on the structure and magnetism \cite{dhar96,flan98,bobev05}, revealing controversial physical properties as well.  Table I briefly compiles the previously reported structure and magnetism of different members of $R$Al$X$.

\begin{table}[b]
\caption{\label{tab:table1}
Crystal structure and magnetism of typical $R$Al$X$ compounds reported in literature. Here NCS and CS denote the noncentrosymmetric LaPtSi type and the centrosymmetric $\alpha$-ThSi$_2$ type structure, respectively. Note that most compounds have a wide homogeneous range and are slightly off-stoichiometric. SG denotes spin glass. }
\begin{ruledtabular}
\begin{tabular}{lcr}
\textrm{$R$Al$X$}&
\textrm{Structure and magnetism}\\
\colrule
CeAlGe & NCS/AFM \cite{hodo18}; CS/AFM \cite{dhar96}; CS/FM \cite{flan98} \\
PrAlGe & NCS/SG \cite{pub19};NCS/FM \cite{meng19} \\
CeAlSi & CS/FM \cite{dhar96}; CS/AFM \cite{flan98}  \\
PrAlSi & CS/FM refs. \cite{bobev05,sharma07} and this work \\
\end{tabular}
\end{ruledtabular}
\end{table}

In this paper, we focus on PrAlSi and report the single crystal synthesis, and detailed investigations of its crystal structure, magnetic, transport and thermodynamic properties. Note that the 4$f^2$ configuration of the Pr$^{3+}$ ion, as confirmed for PrAlSi, is at the center of various exotic physical properties like magnetic ordering, metal-insulator transition and heavy fermion behavior of a variety of Pr-based intermetallics \cite{maple07}. In order to investigate the 4$f$ contributions to these quantities, the corresponding nonmagnetic compound LaAlSi without $f$ electron was also prepared and studied. Structure analysis of PrAlSi by x-ray diffraction reveals a centrosymmetric lattice, with random occupation of Al and Si atoms on the same crystallographic site, see Table II.

By performing dc and ac magnetic measurements, we are able to construct a complicated magnetic phase diagram for PrAlSi, which includes not only a FM phase below $T_C$, but two spin-glass-like reentrant magnetic phase transitions at $T_{\rm M1}$ and $T_{\rm M2}$ that are below $T_C$. Electrical resistivity measurements show large, nonsaturating magnetoresistance in magnetic field, once the spin-glassy phases are suppressed. Specific heat measurements indicate a non-Kramers doublet ground state and a relatively small overall splitting (less than 100 K) of the Pr$^{3+}$ multiplets due to the crystal electric field (CEF). The latter effect largely enhances the magnetic entropy associated with the FM order. Hall-effect measurements reveal a large anomalous Hall conductivity in the FM state, amounting to 2000 $\Omega ^{-1}$cm$^{-1}$ at $T$ $<$ $T_C$. Finally, we introduce the Shubnikov-de Hass (SdH) oscillations of $\rho(B)$ observed from slightly above $T_{\rm C}$ down to $T$ = 2 K for PrAlSi. Intriguingly, the dominating oscillation frequency is strongly temperature dependent, changing from $F$ = 18 T at 25 K to $F$ = 33 T at 2 K. By contrast, no SdH oscillations can be observed for the nonmagnetic LaAlSi in the same temperature window. These results signify a small but largely expanded Fermi surface in PrAlSi upon cooling into the FM phase, suggestive of a significant coupling of the local magnetism of Pr$^{3+}$ ions and the small Fermi pocket of $s$ and $p$ electrons \cite{chang18}.

\section{experimental details}
Single crystals of PrAlSi and LaAlSi were grown from high-temperature self flux using molten Al as solvent \cite{bobev05}.  High purity chunks of cerium/lanthanum, silicon, and aluminum were loaded into an alumina crucible in the mole ratio 1:1:10, and further sealed in a quartz tube under high vacuum. The loaded quartz tube was slowly heated up to 1150$^{\rm o}$C in 12 h, held at that temperature for 2 h in order to ensure an enough melting. It was then cooled down to 750$^{\rm o}$C in 100 h and dwells for 2 days. The excess Al was then removed by centrifuging at 750$^{\rm o}$C at the end of the growing process. Large mirror-like plates of single crystal PrAlSi and LaAlSi, see inset of Fig. 1, were obtained. Trace of residual aluminium on the surface of the obtained single crystals was removed in dilute solution of NaOH.

To identify the crystal structure of PrAlSi, single crystal x-ray diffraction spectrum was collected at room temperature by employing the Bruker D8 Venture diffractometer. The crystal structure was refined by full-matrix least-squares fitting on the structure factor $F^2$ using the SHELXL-2014/7 program. The electrical resistivity and Hall effect were measured in the physical property measurement system (PPMS, Quantum Design) between 2 K and room temperature, using a sample of typical dimension 0.3$\times$1$\times$3 mm$^3$. The electrical current was applied within the tetragonal basal plane. Five-contact technique was employed for the Hall-effect measurement, where a weak field-even term due to longitudinal magnetoresistance was eliminated by scanning field in negative and positive fields.  The dc magnetic susceptibility and magnetization measurements were carried out in the MPMS-SQUID magnetometer and the ac susceptibility was measured in various applied dc bias fields by employing the ac susceptometer equipped on PPMS. The specific heat measurements for both PrAlSi and LaAlSi were performed by a thermal-relaxation method.

\begin{table}
\caption{\label{tab:table1}
Refinement results of atomic information obtained for PrAlSi at room temperature. $U_{eq}$ is the thermal displacement factor in unit \AA$^2$.}
\begin{ruledtabular}

\begin{tabular}{lcccccr}
\multicolumn{7}{l}{Refined chemical formula: PrAl$_{1.13}$Si$_{0.87}$} \\
\multicolumn{7}{l}{Crystal structure: Tetragonal $\alpha$-ThSi$_2$}\\
\multicolumn{7}{l}{Space group: I4$_1$/amd (No. 141)}\\
\multicolumn{7}{l}{Lattice constants: $a$ = $b$ = 4.2255 \AA, $c$ = 14.534 \AA}\\
\colrule
\textrm{atom}&
\textrm{wyck.}&
\textrm{$x$}&
\textrm{$y$}&
\textrm{$z$} &
\text{$U_{eq}$} &
\textrm{o.p.}\\

\colrule
Pr & 4a & 0.5  & 0.75 & 0.375 & 0.0063 & 1\\
Al & 8e & 0.5 & 0.25 & 0.2082 & 0.008& $\sim$ 0.6\\
Si & 8e &  0.5 & 0.25  &  0.2082 & 0.008 & $\sim$ 0.4

\end{tabular}
\end{ruledtabular}
\end{table}

\begin{figure}
\includegraphics[width=0.9\linewidth]{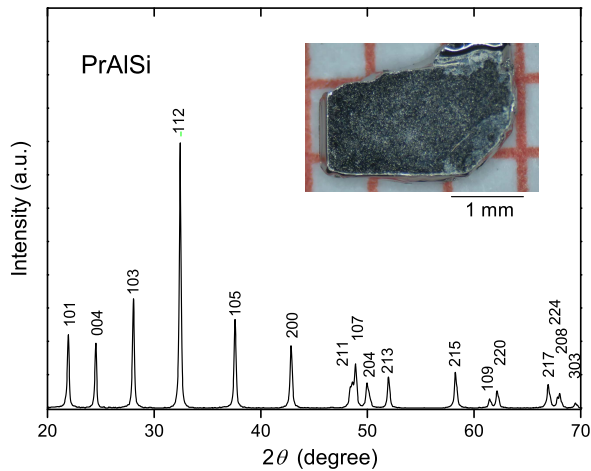}
\caption{Powder x-ray diffraction spectrum of PrAlSi, with all Bragg peaks properly indexed. Note that both the $\alpha$-ThSi$_2$ and LaPtSi-type structures give the same index. Inset shows a photo image of the PrAlSi sample used for transport measurements, with $c$ axis perpendicular to the as-grown plate.
\label{xrd}}
\end{figure}

\section{Experimental results and discussion}

\subsection{Crystal structure}

As already mentioned, two different but closely related crystal structures of LaPtSi and $\alpha$-ThSi$_2$ types are known for $R$Al$X$.
The two structures differ in the sense that Al and $X$ atoms can be either ordered and occupy two different Wyckoff $4a$ sites or, alternatively, disordered and occupy the same $8e$ site; see ref. \cite{hodo18}. Powder x-ray diffraction spectrum (Fig. 1) shows good agreement with other compounds of this family \cite{hodo18,pub19} and reveals no secondary phase. But it can barely provide information on the difference of the two related structures.

Our single crystal x-ray diffraction analysis on PrAlSi supports a centrosymmetric symmetry with space group $I4_1/amd$. As shown in Tab. II, the best refinement indicates that Al and Si atoms occupy the same 8$e$ site randomly, with a composition PrAl$_{1.13}$Si$_{0.87}$ that is in reasonable agreement with that obtained from energy-dispersive x-ray spectrum. The consequently determined lattice constants are $a$ = 4.2255 \AA, and $c$ = 14.534 \AA. Excess of Al and deficiency of Si in $R$AlSi have so far been frequently detected in flux-grown samples from molten Al (ref. \cite{bobev05}), which again points to the disordered nature of the Al and Si atoms in $R$AlSi. This situation is consistent with the previous report \cite{bobev05} of a large homogenous range of PrAl$_x$Si$_{2-x}$ within the $\alpha$-ThSi$_2$ structure. We note that, in the presence of space-inversion symmetry, Weyl nodes cannot emerge naturally unless spin degeneracy is lifted through, e.g. breaking of time-reversal symmetry by applying external field or forming FM order.

\begin{figure*}
\includegraphics[width=0.7\linewidth]{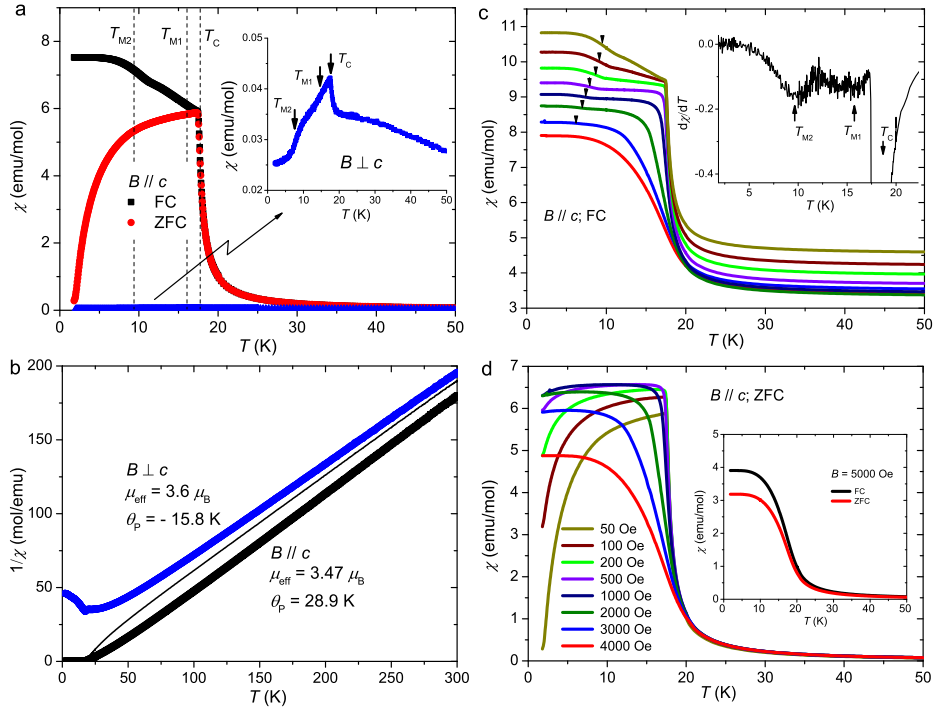}
\caption{(a) The dc magnetic susceptibility $\chi(T)$ measured in ZFC and FC modes in an external field $B$ = 50 Oe.  When $B$$\parallel$$c$, except for the FM transition at $T_C$ = 17.8 K, two subsequent weak anomalies at lower temperatures $T_{\rm M1}$ and $T_{\rm M2}$, are also observed. These anomalies can be better resolved in the derivative d$\chi$/d$T$ shown in the inset of penal (c). Inset: A low-temperature close-up of $\chi(T)$ measured with $B$$\perp$$c$. (b) The inverse susceptibility reveals linear $T$ dependence from room temperature down to $T$ $\approx$ 100 K, with a positive intercept at $\theta_p$ = 28.9 K and a negative one at $\theta_p$ = $-$15.8 K for $B$$\parallel$$c$ and $B$$\perp$$c$, respectively. The obtained effective moments are close to that of the free Pr$^{3+}$ ion. The solid line represents a polycrystalline average $1/\chi_{\rm poly}$. Panels (c) and (d) show FC and ZFC susceptibility $\chi(T)$, respectively, measured in varying dc fields aligned along $c$. Inset of (c) displays the derivative d$\chi$/d$T$ of the FC $\chi(T)$ curve measured in $B$ = 50 Oe. Inset of (d) displays the ZFC and FC $\chi(T)$ results obtained for $B$ = 5000 Oe.
\label{MTPr}}
\end{figure*}

\begin{figure*}
\includegraphics[width=0.99\linewidth]{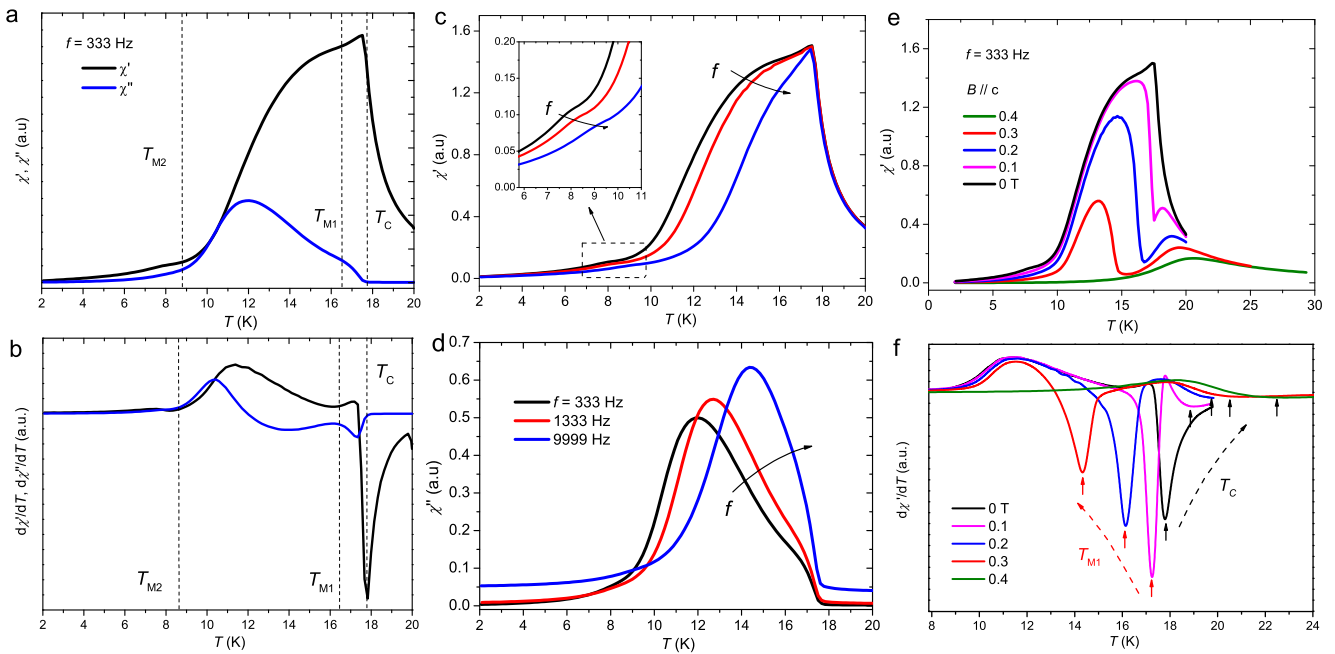}
\caption{The ac magnetic susceptibility of PrAlSi. (a) Real ($\chi'$) and imaginary ($\chi''$) components measured in an ac field of 10 Oe and $f$ = 333 Hz. The corresponding derivatives d$\chi'$/d$T$ and d$\chi''$/d$T$ are shown in panel (b). The FM transition at $T_C$ and the two subsequent transitions at $T_{\rm M1}$ and $T_{\rm M2}$ are indicated by vertical lines, which mark the temperatures where d$\chi'$/d$T$ assumes a minimum. (c, d) Temperature dependence of $\chi'(T)$ (c) and $\chi''(T)$ (d) measured in various frequencies $f$ in zero dc bias field. While $T_C$ is robust to the change of ac field frequency, the anomalies corresponding to $T_{\rm M1}$ and $T_{\rm M2}$ shift to higher temperatures with increasing $f$. (e, f) Temperature dependence of $\chi'(T)$ (e) and d$\chi'(T)$/d$T$ (f) in varying dc bias fields. It can be clearly seen from panel (f) that $T_C$ increases, whereas $T_{\rm M1}$ decreases, with increasing dc field.
\label{MTPr}}
\end{figure*}

\begin{figure}
\includegraphics[width=1\linewidth]{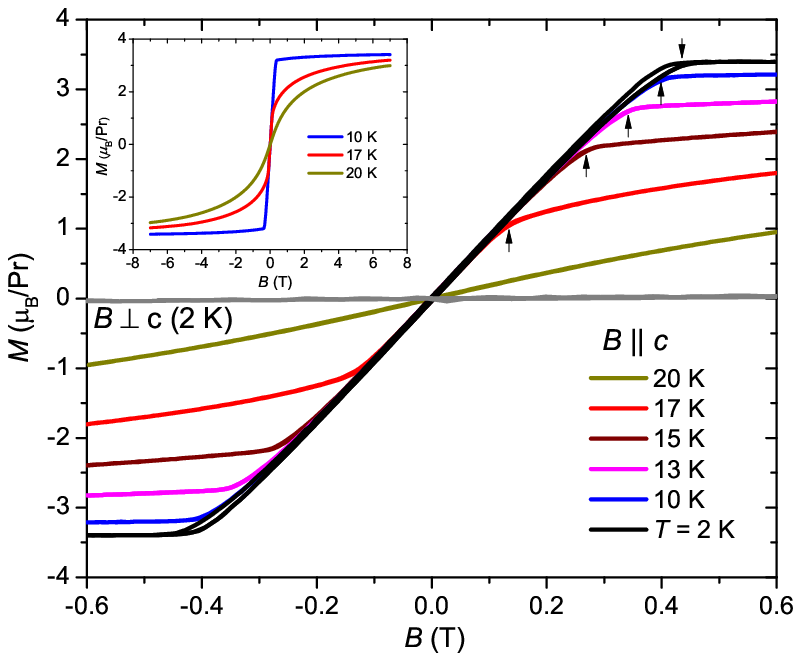}
\caption{Magnetization as a function of field ($B$$\parallel$$c$) for selected temperatures below and around $T_C$.
Arrows indicate critical field $B_c$ at which the low-field magnetic glassy phase changes to the high-field FM phase. Unlike typical ferromagnet, only a very weak hysteresis is visible in fields slightly below $B_c$ in the $M(B)$ curve measured at $T$ =  2 K. $M(B)$ for $B$$\perp$$c$ ($T$ = 2 K) is also shown in order to illustrate the large magnetic anisotropy. Inset shows $M(B)$ measured up to a higher field of 7 T for $T$ = 10, 17 and 20 K.
\label{kappa}}
\end{figure}

\subsection{Magnetic Properties}
Figure 2 summarizes the experimental results of dc magnetic susceptibility measurements. In a small external magnetic field ($B$ = 50 Oe) aligned along $c$ axis, a drastic increase of $\chi(T)$ upon cooling, indicative of a FM transition, is observed in both the field-cooled (FC) and zero-field-cooled (ZFC) $\chi(T)$ curves, see Fig. 2(a). We define the Curie temperature $T_C$ = 17.8 K from the sharp minimum revealed in d$\chi$/d$T$($T$), as shown in Fig. 2(c) inset. The large ratio $\chi_{\parallel c}$/$\chi_{\perp c}$ $\approx$ 150 at $T$ = $T_C$ clearly manifests an Ising-type magnetic anisotropy. Our observation of FM transition at 17.8 K is consistent with a brief report on this compound published previously \cite{bobev05}. The samples employed in the latter work were PrAl$_{1+x}$Si$_{1-x}$ with $x$ = 0.15 and 0.19, similar in composition to our sample and hinting at a relation between stoichiometry and magnetic properties. At $T$ $<$ $T_C$, there appear two more weak anomalies at $T_{M1}$ $\simeq$ 16.5 K and $T_{M2}$ $\simeq$ 9 K, as indicated by arrows in Fig. 2(a), which can be better recognized in the d$\chi(T)$/d$T$ curves (Fig. 2(c) inset) and by ac susceptibility to be shown below. Correspondingly, the FC and ZFC $\chi(T)$ bifurcate significantly below $T_C$, indicative of spin-glass behavior below $T_{\rm M1}$ and $T_{\rm M2}$. Occurrence of spin-glass phase below a FM/AFM transition is usually referred to as reentrant spin glass or cluster spin glass \cite{belik07}. It has been frequently observed in various fragile magnets with certain type of magnetic frustration or atomic disorder, see, for example, Eu$_x$Sr$_{1-x}$S (ref. \cite{maletta80}) and  Mn$_3$Sn (ref. \cite{feng06}). Similar thermal irreversibility of $\chi(T)$ at $T$ $<$ $T_C$ has also been observed for PrAlGe \cite{pub19,meng19}. All the three phase transitions of PrAlSi were also confirmed in $\chi(T)$ measured with $B$$\perp$$c$, see Fig. 2(a) inset.

Shown in Fig. 2(b) are the $T$-dependent inverse susceptibilities $1/\chi$ for both $B$$\parallel$$c$ and $B$$\perp$$c$. From the linear variation of the Curie-Weiss behavior at $T$ $>$ 100 K, the effective magnetic moment $\mu_{\rm eff}$ are estimated to be 3.47 and 3.6 $\mu_{B}$, respectively, close to the free moment of the trivalent Pr ion, 3.58 $\mu_B$. On the other hand, the paramagnetic Weiss temperature $\theta_{p}$ is strongly anisotropic, being 28.9 K ($B$$\parallel$$c$) and $-$15.8 K ($B$$\perp$$c$). The very different values of $\theta_{p}$ point to a competition of out-of-plane FM and in-plane AFM interactions in PrAlSi. This feature is further reflected by the inverse susceptibility of the calculated polycrystalline average $\chi_{\rm poly}$ = (2$\chi_{\perp c}$+$\chi_{\parallel c}$)/3 (solid line) shown in Fig. 2(b). The small value of $\theta_{p}$ = 4 K for $\chi_{\rm poly}(T)$ implies that the FM transition at $T_C$ = 17.8 K is fragile.

To shed light on the nature of the multiple magnetic phases, in Figs. 2(c) and 2(d) we show the $\chi(T)$ curves recorded in different external fields ($B$$\parallel$$c$) in FC and ZFC conditions, respectively. What can be readily recognized is a smooth decrease of $T_{\rm M2}$ upon increasing field in the FC $\chi(T)$ measurements, see Fig. 2(c), where the variations of $T_C$ and $T_{\rm M1}$ cannot be resolved due to their clossness in temperature. Meanwhile, the drastic decrease of the ZFC $\chi(T)$ values below $T_C$ is gradually suppressed upon increasing $B$, see Fig. 2(d). In Fig. 2(d) inset, we display the ZFC and FC $\chi(T)$ curves measured in a relatively high field, $B$ = 0.5 T. There, the two $\chi(T)$ curves become qualitatively similar, with no additional anomalies at $T$ $<$ $T_C$.

The reentrant spin-glass phases can be further probed by ac susceptibility. In Fig. 3(a) we show the real ($\chi'$) and imaginary ($\chi''$) components of the ac susceptibility measured as a function of temperature in an ac field of 10 Oe and frequency $f$ = 333 Hz. The curve of $\chi'(T)$ shows a steep increase at $T_C$, defined by the position of the sharp minimum in d$\chi'$/d$T$, see Fig. 3(b). Unlike typical ferromagnets, $\chi'(T)$ does not drop quickly at $T$ $<$ $T_C$  (ref. \cite{balan13}). Instead, broad humps appear in $\chi'(T)$ due to the weak phase transitions at $T_{\rm M1}$ and $T_{\rm M2}$, as better illustrated in the temperature derivatives of $\chi'(T)$ and $\chi''(T)$, see Fig. 3(b). Accompanying the FM transition, enhanced values of $\chi''(T)$ are also observed, see Fig. 3(a). Moreover, $\chi''(T)$ shows a shoulder at $T_{\rm M1}$ and its values are further enhanced with cooling. These indicate that the magnetic phase below $T_{\rm M1}$ is unlikely to be a (canted) AFM phase, where $\chi''(T)$ is expected to vanish \cite{balan13}. Similar shoulder in $\chi''(T)$, albeit weak, is also observed at $T_{\rm M2}$.

As seen in Figs. 3(c) and 3(d), the positions of $T_{\rm M1}$ and $T_{\rm M2}$ shift to higher temperatures as the frequency of ac field increases. This clearly demonstrates spin-glass freezing within the two subsequent magnetic phases. This notion is also consistent with the strong bifurcation of the ZFC and FC $\chi(T)$ curves observed in low fields, see Fig. 2(a).  Figs. 3(e) and 3(f) show the real component $\chi'(T)$ and the corresponding temperature derivative d$\chi'(T)$/d$T$ measured in various external dc bias fields, respectively. With applying dc field, the magnetic transitions $T_C$ and $T_{\rm M1}$ separates from each other increasingly: The former shifts to higher temperature, as expected for a second-order FM transition; the latter to lower temperature and is suppressed already in a small field of 0.4 T.

Figure 4 displays the isothermal magnetization $M(B)$ for selected temperatures. At $T$ $<$ $T_C$ and when $B$$\parallel$$c$, one observes a linear increase of $M(B)$ until a critical field $B_c$ from which it flattens out. At $T$ = 2 K, $M(B)$ saturates to 3.4 $\mu_B$/Pr at $B_c$ $\approx$ 0.43 T, a value close to the full moment of free Pr$^{3+}$ ion ($gJ$ = 3.2 $\mu_B$). Magnetization measured at higher temperatures $T$ = 10, 17 and 20 K show that a similar saturated moment is attained at higher fields (inset of Fig. 4). Polarization of the full Pr$^{3+}$ moment indicates that the splitting of its ninefold multiplets ($J$ = 4) in CEF, as well as  the Kondo screening effect (if any) are rather weak as compared to the FM interaction. The critical field $B_c$, marked by arrows, decreases all the way with increasing temperature and vanishes at $T$ $\approx$ $T_C$. Characterizing the magnetic glassy state in field below $B_c$ and differing from typical ferromagnet, only a weak hysteresis can be discerned in a small field interval slightly below $B_c$, see $M(B)$ for $T$ = 2 K. These features indicate that the step-like magnetization at $B_c$ does not represent spontaneous FM polarization, but field-induced transition from spin glass to FM phase. The values of $B_c$ obtained for different temperatures will be used to construct the magnetic phase diagram. Reversely, in the case of $B$$\perp$$c$, $M(B)$ measured at $T$ = 2 K (gray line) reveals negligible change up to 7 T, the largest magnetic field accessible in this work.

\begin{figure}
\includegraphics[width=0.96\linewidth]{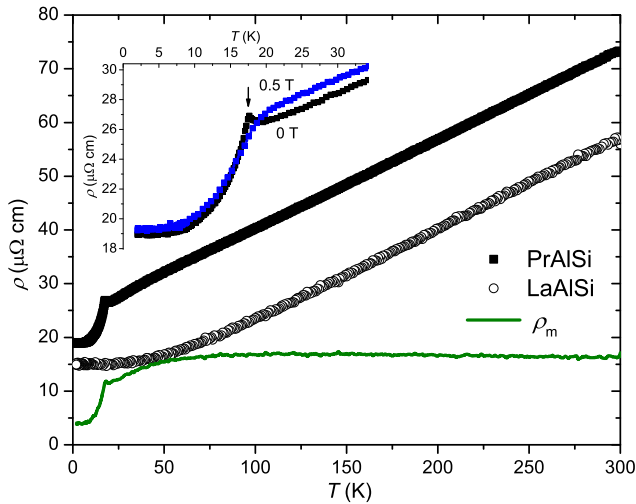}
\caption{Temperature-dependent resistivity $\rho(T)$ of PrAlSi and its nonmagnetic analogue LaAlSi. The magnetic contribution $\rho_{m}(T)$ is estimated by subtracting the $\rho(T)$ values of the latter compound from the former. Inset shows an enlarged view of the low-$T$ resistivity, where a clear cusp can be seen at $T$ $\approx$ $T_C$ for the zero-field measurement. Application of a small magnetic field ($B$ = 0.5 T) smears out this feature.
\label{kappa}}
\end{figure}

\subsection{Transport Properties}

\begin{figure*}
\includegraphics[width=1\linewidth]{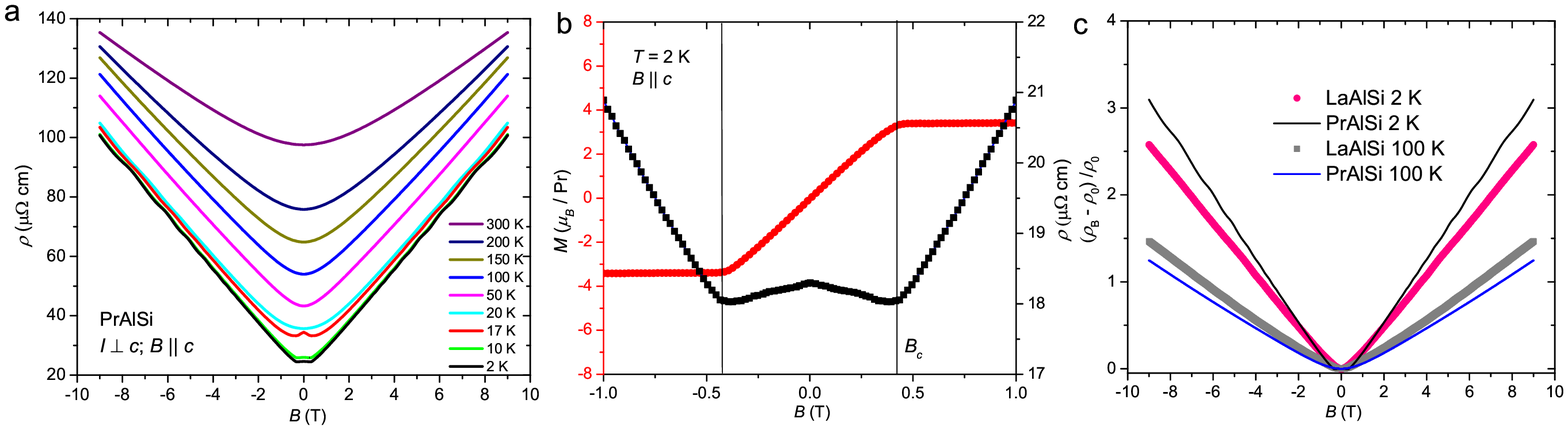}
\caption{(a) Isothermal magnetoresistivity $\rho(B)$ of PrAlSi measured at varying temperatures between $T$ = 2 and 300 K. Below $T_{C}$ = 17.8 K and in low magnetic fields $B$ $<$ $B_c$, MR is only weakly field-dependent and slightly negative, see penal (b) for the closeup of $\rho(B)$ for $T$ = 2 K. At $B$ $>$ $B_c$, MR becomes positive and reveals a nonsaturating, nearly $B$-linear behavior up to 9 T. Magnetic quantum oscillations were also observed below $\sim$25 K for PrAlSi. (b) Comparison of low-field $\rho(B)$ and $M(B)$ curves measured at $T$ = 2 K.  The critical field $B_c$ observed in both quantities agrees very well.  (c) Comparison of the magnetoresistivity ratio ($\rho_B$$-$$\rho_0$)/$\rho_0$ between PrAlSi and LaAlSi at two selected temperatures, $T$ = 2 K and 100 K. While the magnitude of MR is similar for the two compounds, only the magnetic compound PrAlSi shows clear SdH oscillations at low temperatures. No SdH oscillations can be confirmed for LaAlSi.
\label{dkappa}}
\end{figure*}

Figure 5 displays the electrical resistivity $\rho(T)$ measured within the basal plane for both PrAlSi and LaAlSi. Due to, presumably, the atomic disorder between Al and Si sites inherent to these compounds, their residual resistivity ratio $\rho_{\rm 300K}$/$\rho_{\rm 2K}$ $\approx$ 3.8 is rather small. This appears to be a feature generic to many $R$Al$X$ family members \cite{hodo18}. From $T$ = 300 K down to approximately 100 K, $\rho(T)$ of both compounds changes quasi-linearly with temperature due to the dominating acoustic phonon scattering. Upon further cooling, a weak upward bending away from linearity can be seen around 50 K for PrAlSi, whereas $\rho(T)$ of the nonmagnetic compound LaAlSi flattens out below this temperature. The magnetic contribution, $\rho_{m}$, estimated by subtracting the resistivity of LaAlSi from that of PrAlSi, reveals a weak temperature dependence at $T$ $>$ $T_C$ except for the broad hump around 50 K. Without clear $-$ln$T$ dependence characteristic of Kondo effect, this broad hump is most probably derived from CEF effect. Consistently, the inverse susceptibility 1/$\chi$ deviates from Curie-Weiss law below $T$ $\approx$ 50 K, too, see Fig. 2(b).

$\rho(T)$ of PrAlSi does not simply drop at $T_C$ as is generally expected for local-moment based metallic ferromagnets. Instead, it starts to increases slightly  about $T_C$ and develops a small cusp (Fig. 5 inset) before decreasing upon further cooling. The cusp at $T_C$ is field sensitive and can be easily suppressed by a small field (e.g., $B$ = 0.5 T) applied along $c$ axis. We found such resistivity behavior is rather common for FM semimetals with low charge-carrier density and has been observed in, for example, EuCuP (ref. \cite{iha19}) and EuB$_6$ (ref. \cite{sullow98}). The unusual increase of resistivity above $T_C$ has been interpreted in two alternative scenarios specific to magnetic semimetals: One relies on the critical spin fluctuations near the magnetic phase transition \cite{kata01} and the other is based on the formation of magnetic polaron \cite{yu05}.

Figure 6(a) displays the isothermal magnetoresistivity (MR) $\rho(B)$ measured in transverse magnetic fields ($I$$\perp$$c$, $B$$\parallel$$c$) for PrAlSi.  At $T$ $<$ $T_{C}$, $\rho(B)$ first weakly decreases with field (namely, a negative change of MR) until the critical field $B_c$, see Fig. 6(b) for a comparison of $\rho(B)$ and $M(B)$ at $T$ = 2 K. At $B$ $>$ $B_c$, where FM phase is recovered, $\rho(B)$ becomes a positive function and increases quasi-linearly up to at least $B$ = 9 T, see Figs. 6(a) and 6(b). The negative change of MR below $B_c$ is due to spin disorder scattering and is a generic feature of spin glass \cite{perez98}. The nonsaturating behavior of $\rho(B)$ above $B_c$ is observed all the way up to room temperature, characteristic of a compensated or topological semimetal. Even at 300 K, the MR ratio defined as ($\rho_{B}$$-$$\rho_{0T}$)/$\rho_{0T}$ is large and amounts to 40\% for $B$ = 9 T, surmounting that of some topological semimetals \cite{ali14}.

At temperatures below about 20 K, SdH quantum oscillations are observed for PrAlSi down to a relatively small field $B$ $\approx$ 3 T (Fig. 6(a)), manifesting the small Fermi pocket of this compound. In Fig. 6(c), we compare the MR ratio of PrAlSi and LaAlSi measured at two representative temperatures, $T$ = 2 and 100 K. Interestingly, no SdH quantum oscillations can be confirmed for LaAlSi in the parameter range down to 2 K and up to 9 T, despite similar sample quality of the two compounds.  These facts suggest that the onset of FM order in PrAlSi has a profound influence on its electronic structure. Consequently, the SdH quantum oscillations and their unusual temperature dependence will be analyzed below as reflecting an emerging Fermi pocket in the ordered state PrAlSi.

\begin{figure*}
\includegraphics[width=1\linewidth]{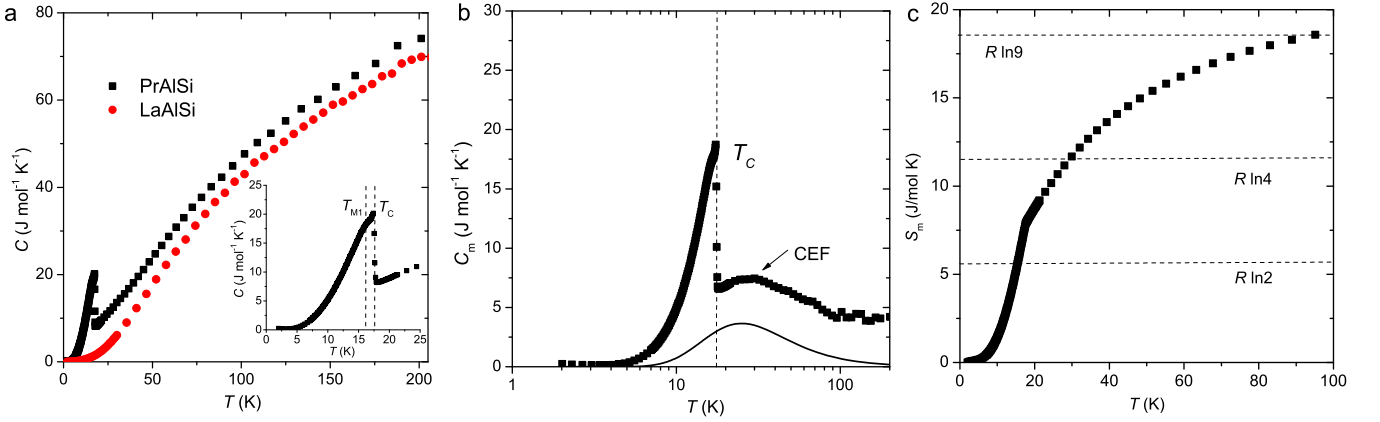}
\caption{(a) Specific heat as a function of temperature for PrAlSi and LaAlSi. In addition to the $\lambda$-type peak observed at $T_C$, a broad should at $T$ $\approx$ $T_{\rm M1}$ is also visible, see inset. (b) Magnetic contribution to the specific heat, $C_{m}$, estimated by subtracting the specific heat of LaAlSi from that of PrAlSi, is plotted as a function of temperature. A Schottky maximum due to CEF splitting of the Pr$^{3+}$ multiplets appears at about 30 K, a temperature rather close to $T_C$. To estimate the overall CEF splitting energy, a solid line calculated simply from a ground state doublet and an excited doublet at 60 K is also shown. (c) Magnetic entropy $S_{m}$, obtained by integrating $C_{m}/T$ with respect to $T$, is shown as a function of temperature.
\label{kappa}}
\end{figure*}

\begin{figure}
\includegraphics[width=0.92\linewidth]{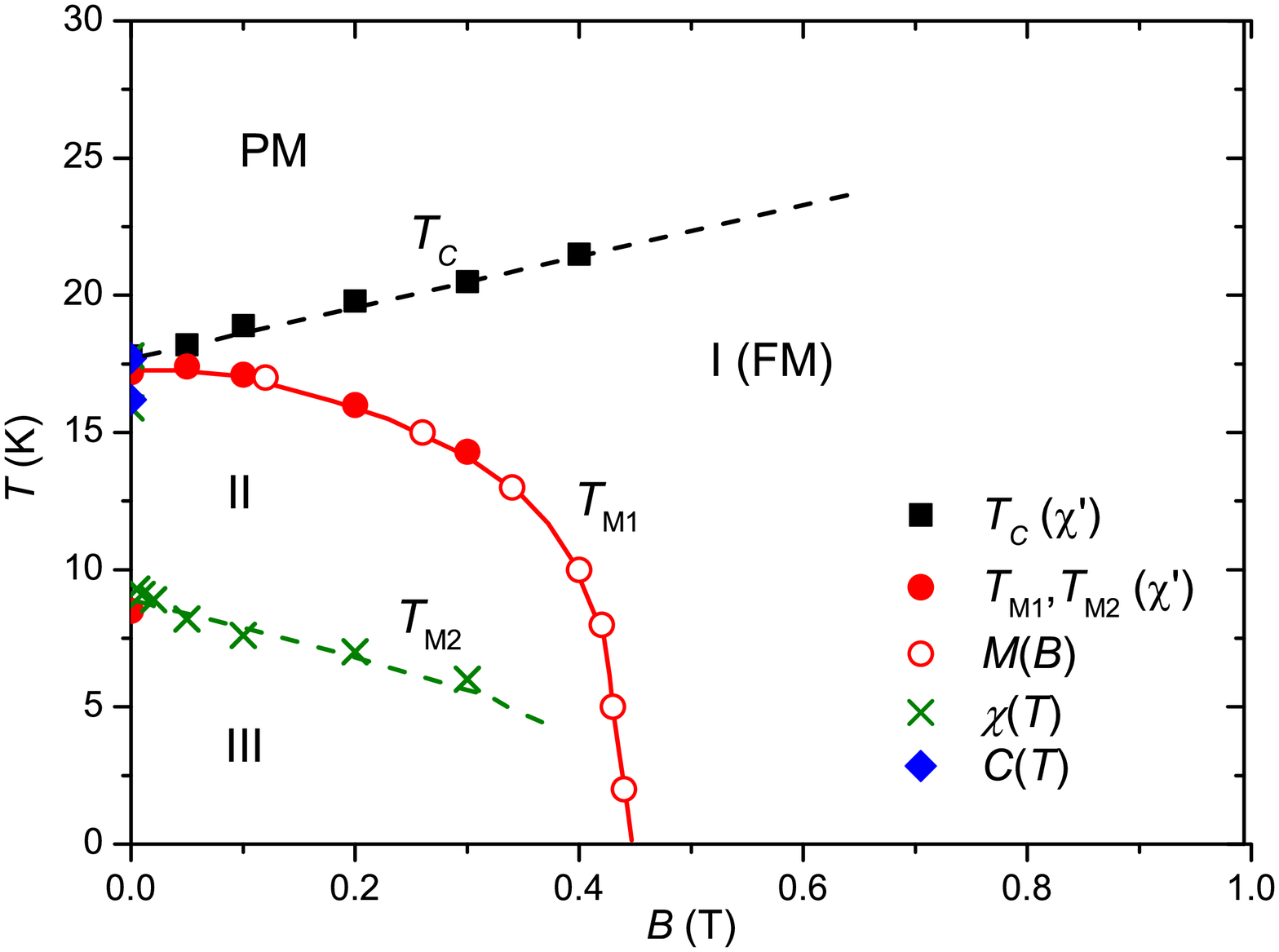}
\caption{Temperature-field magnetic phase diagram of PrAlSi. The FM transition temperature $T_{C}$ slightly shifts upward in magnetic field, yielding an extended FM phase in field (I). By contrast, the two broad phase transitions at $T_{\rm M1}$ and $T_{\rm M2}$ are gradually suppressed by the application of magnetic field, leading to two closed magnetic glassy phases in the low-field low-temperature corner, labeled as II and III.
\label{kappa}}
\end{figure}

\subsection{Specific heat}

Figure 7(a) shows the temperature-dependent specific heat $C(T)$ for both PrAlSi and LaAlSi. A sharp $\lambda$-type peak at $T_C$ = 17.8 K can be observed for PrAlSi. Right below $T_C$, the tail of the $\lambda$-type anomaly is interrupted by a weak shoulder at $T$ $\approx$ 16 K, see Fig. 7(a) inset. The latter value agrees reasonably to $T_{\rm M1}$ determined by susceptibility (Figs. 2-3), indicating a bulk nature of the weak phase transition at $T_{\rm M1}$. No feature at $T_{\rm M2}$ can be discerned from $C(T)$, though it can be rather clearly observed in susceptibility, see Figs. 2(a) and 2(c).

In Fig. 7(b), the magnetic contribution to specific heat $C_{m}(T)$, obtained by subtracting $C(T)$ of the nonmagnetic reference LaAlSi from that of PrAlSi, is displayed. Remarkably, $C_{m}(T)$ reveals a broad maximum at $T$ $\sim$ 30 K, on which sits the peak of the FM transition at $T_C$ = 17.8 K. Note that the two features are rather close in temperature. The broad $C_{m}(T)$ maximum is a Schottky contribution arising from the CEF splitting of the Pr$^{3+}$ multiplets.  To shed light on the CEF scheme, the magnetic entropy estimated by integrating the values of $C_{m}/T$ with respect to $T$ is shown in Fig. 7(c). Here, the small portion of the magnetic entropy below $T$ = 2 K, the lowest temperature of out measurements, is ignored. The estimated entropy $S_m$ at $T_C$ amounts to 7.9 J/mol K, substantially larger than $R$ln2 (5.76 J/mol K), i.e., the magnetic entropy associated with a doublet ground state, but much smaller than $R$ln4. Apparently, the magnetic entropy released below $T_C$ is largely influenced by the broad Schottky contribution. If the latter is smoothly extrapolated to below $T_C$ and subtracted from $C_m(T)$, the estimated magnetic entropy at $T_C$ matches reasonably well with $R$ln2 (not shown). The ground state of PrAlSi is therefore most likely a non-Kramers magnetic doublet, considering the 4$f^2$ configuration of Pr$^{3+}$ ion.

Within the $\alpha$-ThSi$_2$-type structure, the Pr$^{3+}$ ions in PrAlSi adopt the $D_{\rm 2d}$ ($-$42m) point symmetry. The corresponding CEF will split the $J$ = 4 multiplets into five singlets and two non-Krammer doublets, as was discussed for PrSi$_2$ \cite{dhar94}. The fact that the full magnetic entropy of the nine-fold multiplets ($R$ln9) is released at 100 K (see Fig. 7(c)) indicates that all the excited CEF states locate within this relatively narrow energy window. In order to provide a further estimate to the overall splitting energy, in Fig. 7(b), we show a simple calculation of the CEF contribution (solid line) based on a ground state doublet and an excited doublet at $\Delta$ = 60 K, see ref. \cite{dhar94} for the calculation procedure. As can be observed, this line qualitatively reproduces the observed $C_{m}(T)$ maximum at $T$ $\sim$ 30 K. Therefore, this $\Delta$ value can be considered a proper energy scale of the degeneracy center of the overall CEF excitations, though details of the CEF scheme are yet to be clarified.

\subsection{Magnetic phase diagram}

The magnetic phase diagram of PrAlSi derived from the aforementioned experiments for $B$$\parallel$$c$ is shown Fig. 8.  As revealed by $\chi'(T)$ measurements performed in varying dc magnetic fields (Figs. 3(e) and 3(f)), $T_C$ shifts slightly upwards with increasing field. Actually, the FM phase transition is still visible in a magnetic field of 5 T at a much higher temperature of $\sim$25 K (ref. \cite{sharma07}). In zero field and below the FM transition temperature, there are two reentrant, weak magnetic transitions at $T_{\rm M1}$ $\simeq$ 16.5 K and $T_{\rm M2}$ $\simeq$ 9 K, below which either FM cluster glass or spin glass is formed. Unlike $T_C$, both $T_{\rm M1}$ and $T_{\rm M2}$ can be gradually suppressed by applying magnetic field, forming two closed magnetic phases below $B_c$($T$) in the low-temperature low-field phase space, denoted as II and III, respectively (Fig. 8). As already mentioned, phases II and III are not canted AFM phases, where the enhanced values of $\chi''(T)$, the significant FC-ZFC susceptibility bifurcation, as well as the frequency dependence of ac susceptibility are not expected.

To understand the complicated phase diagram with fragile FM and reentrant magnetic glassy phases, one apparent approach is to consider the atomic disorder and the competing FM and AFM interactions inherent to this compound, as has been discussed for Mn$_3$Sn in ref. \cite{feng06}. For another possibility, we consider the Ruderman-Kittel-Kasuya-Yosida (RKKY) indirect magnetic interaction mediated by conduction electron in a low charge-carrier concentration semimetal. Here we mention a similar case of EuB$_6$, a well-known FM semimetal, where complicated magnetic phase transitions are also observed below $T_C$ (ref. \cite{sullow98}). The low charge-carrier density or, in other words, the small Fermi energy $\epsilon_F$ comparable to the energy scale of the RKKY magnetic interaction, is believed to play a dominating role in forming the complicated phase diagram \cite{ya01}. Theoretically, the magnetic order in such a semimetal is expected to become unstable due to the Coulomb interaction of the charge fluctuations in a small Fermi pocket \cite{ya01}.

\subsection{Normal and anomalous Hall effect}

\begin{figure*}
\includegraphics[width=0.99\linewidth]{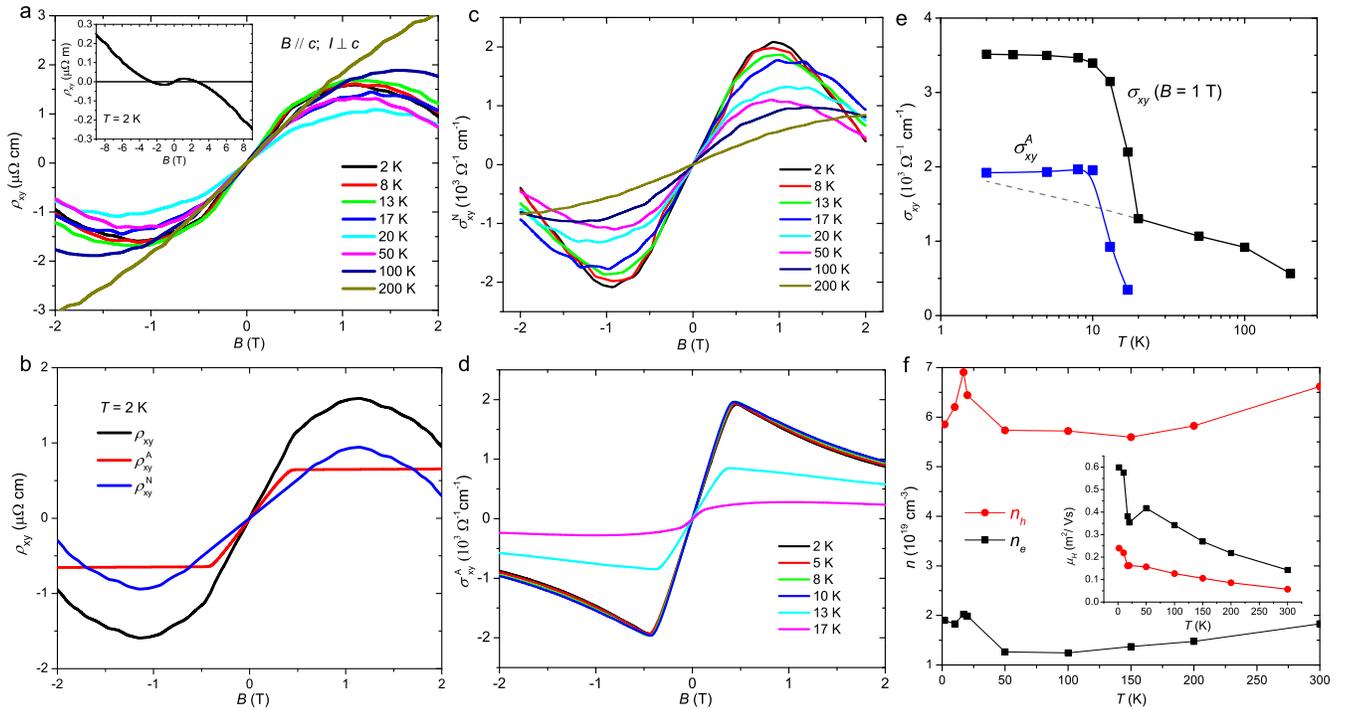}
\caption{(a) The Hall resistivity $\rho_{xy}(B)$, measured with $I$$\perp$$c$ and $B$$\parallel$$c$, is shown for a field window $|B|$ $<$ 2 T for varying temperatures. Inset: $\rho_{xy}(B)$ measured at $T$ = 2 K in a large field range up to 9 T.  (b) The measured $\rho_{xy}(B)$ for 2 K is decomposed to a normal and an anomalous part, assuming $\rho_{xy}$ = $\rho^N_{xy}$ + $\rho^A_{xy}$. Among them, the anomalous component $\rho^A_{xy}$ scales to the magnetization $M(B)$ (see Fig. 4). (c, d) Normal and anomalous Hall conductivities, $\sigma^N_{xy}(B)$ and $\sigma^A_{xy}(B)$, calculated from $\rho^N_{xy}(B)$, $\rho^A_{xy}(B)$ and the as-measured resistivity for varying temperatures. (e) Anomalous Hall conductivity $\sigma^{A}_{xy}(T)$ read off at $B$ = $B_c$ from $\sigma^A_{xy}(B)$ shown in panel (d), and the total Hall conductivity read off at $B$ = 1 T. The dashed line is an extrapolation of the latter value from above $T_C$. (f) Carrier concentration $n$ and Hall mobility $\mu_H$ (inset) as calculated based on the two-band analysis on $\sigma^{\rm N}_{xy}(B)$, cf. panel (c).
\label{dkappa}}
\end{figure*}

The large, intrinsic AHE in FM and topological materials is being intensively revisited in term of the enhanced Berry curvature of occupied electronic states \cite{chen14,burkov15,enke18,naga10}. This issue is of significant interest when ferromagnetism and nontrivial band topology coexist, as realized in FM Weyl semimetals \cite{burkov15}. Indeed, in the FM Weyl semimetal candidate Co$_3$Sn$_2$S$_2$, a sizable anomalous Hall conductivity amounting to above 1100 $\Omega ^{-1}$cm$^{-1}$  has been observed in the ordered phase \cite{enke18,qwang18}. Such effect is attracting increasing attention also because of its potential application in low-energy consumption spintronics \cite{smej17}.

In Fig. 9, we compile our experimental results of the Hall-effect measurements performed for PrAlSi. As seen in Fig. 9(a), the Hall resistivity $\rho_{xy}(B)$ is significantly nonlinear with magnetic field in the temperature ranges both below and above $T_C$. This nonlinearity is caused by multi-band contributions and will be analyzed along this line below. In addition, at $T$ $<$ $T_{C}$ and $B$ $<$ $B_c$, a linear-in-$B$ contribution to $\rho_{xy}(B)$, resembling the corresponding $M(B)$ curve (Fig. 4), can be recognized. Empirically, the Hall resistivity of a ferromagnet can be expressed as $\rho_{xy}(B)$ = $\rho_{xy}^N$ + $\rho_{xy}^A$, where the former is the normal Hall resistivity scaling to $B$ and the latter anomalous one scaling to magnetization, i.e., $\rho_{xy}^A$ = $R_{S}$$M$ ($R_{S}$ is the anomalous Hall coefficient). Based on this description, we can separate the experimental values of $\rho_{xy}(B)$ at $T$ $<$ $T_C$ into $\rho_{xy}^N(B)$  and $\rho_{xy}^A(B)$, as demonstrated in Fig. 9(b) for $T$ = 2 K. Because the low-field reentrant magnetic phases (II and III) show magnetic glassy behaviors, no apparent hysteresis loop can be detected in the Hall resistivity $\rho_{xy}(B)$, different to the FM semimetal Co$_3$Sn$_2$S$_2$ \cite{enke18,qwang18}.

In Fig. 9(c) and 9(d), we show the isothermal normal Hall conductivity $\sigma^N_{xy}(B)$ and the anomalous counterpart $\sigma^A_{xy}(B)$, calculated from the separated $\rho_{xy}^N$ and $\rho_{xy}^A$, as well as the as-measured values of $\rho$ and $\rho_{xy}$,
\begin{equation}
\sigma^N_{xy} = \frac{\rho^N_{xy}}{\rho^2 + \rho_{xy}^2}; \, \sigma^A_{xy} = \frac{\rho^A_{xy}}{\rho^2 + \rho_{xy}^2}.
\end{equation}
One can immediately see that upon entering the ordered phase below $T_C$, while $\sigma_{xy}^N(B)$ does not show significant change with temperature, large values of $\sigma^A_{xy}(B)$ emerge rapidly at $T$ $<$ $T_C$, see Figs. 9(c) and 9(d).

Fig. 9(e) displays the temperature dependence of the anomalous Hall conductivity $\sigma^{A}_{xy}$, which is read off at $B$ = $B_c$ at various temperatures (Fig. 9(d)). For comparison, the as-calculated total Hall conductivity $\sigma_{xy}(B)$ read off at $B$ = 1 T is also shown. Markedly, large values of $\sigma^{A}_{xy}(T)$ rapidly develops at $T$ $<$ $T_C$; similar trend can also be seen in the total Hall conductivity $\sigma_{xy}(T)$ as well.  On the other hand, the normal Hall coefficient $\sigma^{N}_{xy}(T)$ changes only smoothly  below $T_C$, as revealed by the dashed line in Fig. 9(e). The observed $\sigma^{A}_{xy}$ $\approx$ 2000 $\Omega^{-1}$cm$^{-1}$ at $T$ $<$ $T_C$ is even larger than the giant AHE observed in Co$_3$Sn$_2$S$_2$ \cite{enke18}. We note, however, that the large $\sigma^{A}_{xy}$ value of PrAlSi is observed at the critical field $B_c$. It is not a spontaneous Hall conductivity because of the spin glassy state in zero field, different to that of Co$_3$Sn$_2$S$_2$.

Assuming a compensated two-band model, the normal Hall conductivity is expressed as
\begin{equation}
\sigma_{xy}^{N} (B) = \frac{n_ee\mu_e^2B}{1+\mu_e^2B^2} + \frac{n_he\mu_h^2B}{1+\mu_h^2B^2}.
\end{equation}
The nonlinear $\sigma_{xy}^{N}(B)$ shown in Fig. 9(c) can be fitted by the two-band model within the field window from $-2$ T to 2 T.  As shown in Fig. 9(f), the estimated carrier concentrations are of the order of 10$^{19}$ cm$^{-3}$, which is much lower than that of the typical metals $\sim$10$^{22}$$-$10$^{23}$ cm$^{-3}$, characterising PrAlSi as a semimetal. The hole-like carriers have a 3$-$5 times higher concentration than the electron-like carriers, in line with the positive initial slopes of $\rho_{xy}(B)$.  On the other hand, the electron-like minority carriers reveal a higher mobility (inset of Fig. 9(f)), which explains the sign changes of Hall resistivity in higher fields, see inset of Fig. 9(a).

\subsection{Shubnikov-de Hass oscillations}

\begin{figure*}
\includegraphics[width=0.7\linewidth]{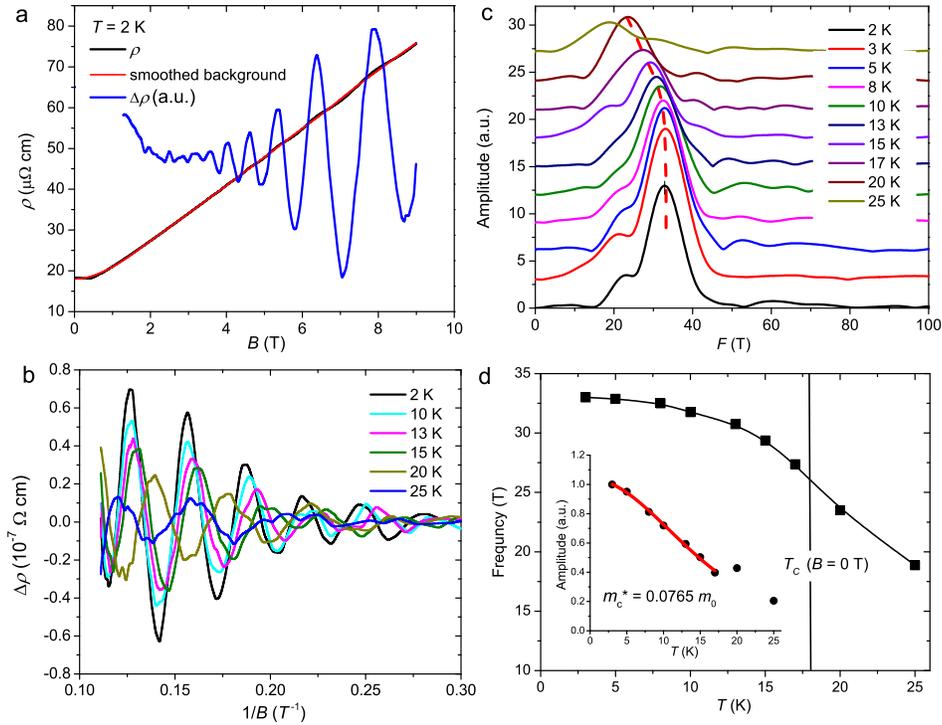}
\caption{(a) Magneto-resistivity $\rho(B)$ measured at $T$ = 2 K in the configuration $I$$\perp$$c$ and $B$$\parallel$$c$ and a smoothed background. Their difference $\delta \rho(B)$ reveals significant SdH oscillations (blue line). $\delta \rho(B)$ at $B$ $<$ 1 T is not shown because of the existence of magnetic phase transitions in this field window, see Fig. 8. (b) SdH oscillations at selected temperatures, shown as $\delta \rho(B)$ vs $B^{-1}$. An apparent change of the oscillation period with temperature can be observed. (c) FFT spectrum of the SdH oscillations for selected temperatures below 25 K. The dominating frequency shifts downwards upon warming and the change is particularly significant at $T$ $\approx$ $T_C$. (d) Temperature dependence of the SdH oscillation frequency. The vertical line indicates the position of $T_C$ = 17.8 K for $B$ = 0 T, and the actual $T_C$ in the field interval of the SdH oscillations is several kelvins higher. Inset: SdH oscillation amplitude as a function of temperature. A theoretical fitting for $T$ $<$ 17 K based on the standard Lifshitz-Kosevich theory yields a small cyclotron effective mass $m^*$ = 0.0765 $m_0$.
\label{kappa}}
\end{figure*}

As already shown in Figs. 6(a) and 6(c), SdH oscillations are observed in the low-temperature $\rho(B)$ curves of PrAlSi, but not LaAlSi. The SdH oscillations at $T$ = 2 K are demonstrated in Fig. 10(a). Assuming a smooth, almost $B$-linear background (red solid line), the oscillatory part $\delta\rho(B)$ can be obtained as the difference between the measured values of $\rho(B)$ and the smooth background. The consequently obtained $\delta \rho(B)$ as a function of $B^{-1}$ is shown in Fig. 10(b) for selected temperatures below 25 K. Clearly, $\delta \rho(B)$ is a periodic function of $B^{-1}$. However, it reveals a significantly $T$-dependent periodicity spacing. The fast Fourier transformation (FFT) analysis of the SdH oscillations reveals one dominating frequency for all the temperatures, see Fig. 10(c). Intriguingly, the oscillation frequency changes from $F$ = 33 T at 2 K to $F$ = 18 T at 25 K. The change of $F$ is not linear in temperature and becomes in particular significant at around $T_C$, see Fig. 10(d). Because the SdH oscillations are absent in the nonmagnetic analogue LaAlSi of similar sample quality (Fig. 6(c)), it is tempting to relate such behaviors in PrAlSi to the onset of FM order. Moreover, as revealed by the FFT spectrum, a shoulder appears at the left-hand side of the dominating frequency for all temperatures, see Fig. 10(c). Given the two-band nature as revealed by Hall measurements (Fig. 9(f)), this shoulder may be an intrinsic feature reflecting the Fermi surface of the (electron-like) minority band.

The SdH oscillation frequency is directly proportional to the extremal cross section $A_F$ of the Fermi surface perpendicular to the magnetic field through the Onsager relation $F$ = ($\hbar$/2$\pi$$e$)$A_F$. Naively, the strong temperature dependence of $F$ observed for PrAlSi indicates a strongly $T$-dependent $A_F$. Such a change of Fermi surface with temperature (by 40\% upon warming from $T$ = 2 K to 25 K) is surprising. Obviously, it is not due to thermal broadening of the Fermi-Dirac distribution, because the Fermi temperature $T_F$ (i.e., $\epsilon_F$/$k_B$, with $\epsilon_F$ $\sim$ 127 meV, see below) is far above the temperature window where the SdH oscillations are observed. Though very rare, such a strongly temperature-dependent Fermi surface of similar extent has been observed for several $f$-electron-based semimetals like CeBiPt. There, a temperature-dependent hybridization of the conduction and $f$ band was argued to be the origin \cite{goll02}. Contrary to the expectation of this scenario, the estimated cyclotron effective mass is actually very small, as will be revealed from the $T$-dependent oscillation amplitude below. Another possibility is that the small Fermi surface is strongly affected by the onset of FM order, i.e., a spin polarization dependent electronic structure. Actually, for the rare-earth based FM semiconductor EuB$_6$, this effect has to be considered in order to interpret its various physical properties like optical conductivity \cite{kunes04,kim08}. Similarly, the SdH oscillation frequency in FM semimetal Sr$_{1-y}$Mn$_{1-z}$Sb$_2$ ($y, z$ $<$ 0.1) was found to be strongly dependent on the saturated moment \cite{liu17}.  The latter scenario appears to be more reasonable for PrAlSi in view of the following facts. First, the SdH oscillations are observed only in the magnetic PrAlSi but not the nonmagnetic LaAlSi. Second, the oscillation frequency changes rapidly in the vicinity of $T_C$. Here, note that $T_C$ is an increasing function of field and may increase from $T_C$ = 17.8 K in zero field up to 30 K in the field range of SdH oscillatoins. This temperature window of $T_C$ reasonably matches the temperature range where $F$ strongly changes, see Fig. 10(d). Actually, the FM order underlies the formation of Weyl semimetal phase in some centrosymmetric semimetals, such as CeSb \cite{cesb}. However, whether the observed Fermi pocket in PrAlSi is topologically nontrivial remains an intriguing question for this compound.

As shown in Fig. 10(d) inset, by fitting the temperature dependence of the SdH oscillation amplitude at $T$ $<$ 17 K to the temperature damping factor $R_T$ = $\alpha$$X$/sinh($\alpha X$) of the Lifshitz-Kosevich theory, where $\alpha$ = 2$\pi^2$$k_B$/$e\hbar$ and $X$ = $m_c^*$$T$/$B$, we have determined the cyclotron effective mass $m_c^*$ $\approx$ 0.0765 $m_0$. Here, $m_0$ is the bare electron mass. The small effective mass and the small Fermi pocket observed for PrAlSi in its ordered state make it easy to probe the quantum oscillations by transport measurements. Note that our fitting has been confined to $T$ $<$ 17 K where the oscillation frequency does not significantly change, see Fig. 10(d) inset. Based on the Fermi wave vector $k_F$ determined from $A_F$ = $\pi$$k_F^2$ and assuming this pocket is dominating in PrAlSi, the Fermi energy $\epsilon_F$ = $\hbar^2$$k_F^2$/$m^*$ can be estimated to be $\sim$ 127 meV. Note that this value is a rough estimate and will be reduced if one considers the minority band of opposite polarity.

\section{Summary and conclusion}

To summarize, we have synthesized single-crystalline samples of PrAlSi and its nonmagnetic reference compound LaAlSi by self flux method. These compounds are of great recent interest because they belong to a large family of $R$Al$X$ which, if $R$ is a magnetic rare-earth ion, has been predicted to form magnetic semimetal with topologically nontrivial electronic structure. Single-crystal analysis indicates that PrAlSi adopts the $\alpha$-ThSi$_2$ type structure with random occupation of Al and Si atoms in the same crystallographic 8$e$ site, therefore retaining the space-inversion symmetry. Combined dc and ac magnetic measurements on PrAlSi have evidenced not only a FM phase transition at $T_{C}$ = 17.8 K, but also two subsequent reentrant phase transitions at lower temperatures $T_{\rm M1}$ and $T_{\rm M2}$, below which magnetic glassy behaviors are observed. Due to the strong single-ion crystal electric field, the magnetic moments of Pr$^{3+}$ ions are forced to align along the $c$ axis, indicating Ising-type anisotropy. The reentrant magnetic glassy phases can be easily suppressed by a small magnetic field of $B_c$$\sim$0.4 T applied along the magnetic easy $c$ axis, see the magnetic phase diagram shown in Fig. 8.  In higher magnetic fields of $B$ $>$ $B_c$, the FM phase is recovered and large, nonsaturating magnetoresistivity appears, too. Moreover, a huge anomalous Hall conductivity of $\sim$2000 $\Omega^{-1}$cm$^{-1}$ is observed in the FM state, after the spin glassy phases are suppressed by field at $B_c$.

The SdH oscillations observed in PrAlSi reveal a dominant frequency of 33 T at $T$ = 2 K. Upon increasing temperature to the vicinity of the FM transition, not only the oscillation amplitude, but also the oscillation frequency changes rapidly and disappears at $T$ $>$ 25 K. By contrast, no SdH oscillations can be observed in LaAlSi in the same parameter ranges of temperature and field. These facts naively hint at a magnetic modulation to the electronic structure of PrAlSi accompanying the onset of the FM order and a consequently emerging small Fermi pocket detected by SdH oscillations. Whether Weyl nodes appear in the time-reversal symmetry breaking FM phase of PrAlSi, as first predicted for pyrochlore iridates \cite{wan11}, remains an interesting question. Furthermore, except for the $\lambda$-type peak at $T_C$, a broad maximum of magnetic specific heat can be observed at around 30 K. The latter feature can be ascribed to the CEF splitting of the ninefold Pr$^{3+}$ multiplets, with a rather small overall energy scale less than 100 K. The magnetic entropy analysis indicates the CEF ground state of PrAlSi is a non-Kramers doublet. Our results show that PrAlSi may provide a new platform where, the small Fermi pocket of likely relativistic fermions is significantly coupled to ferromagnetism. At last, we mention that PrAlSi also appears suited for investigating the RKKY-type magnetic coupling mediated by conduction electrons in a low charge-carrier concentration semimetal. Related to this, whether electronic correlations originating from Kondo physics are weakly involved in PrAlSi deserves future attention.

\section{Acknowledgement}

The authors thank the fruitful discussions with F. Steglich, Q. Si, Y-f. Yang, and J.L. Luo. This work was supported by the MOST of China (grant no. 2017YFA0303100) and the National Science Foundation of China (grant nos. 11774404, 11474332).

\end{document}